\documentclass[twocolumn, amssymb, amsmath, nobibnotes, nofootinbib, aps, prl,showpacs]{revtex4-1}
\usepackage[dvips]{graphicx}
\begin{document}
\title{Failure to achieve the $J_{eff}$~=~0 state even in nearly isolated Ir$^{5+}$ in Sr$_3$NaIrO$_6$: are iridates enough for realizing true $j$-$j$ coupling?}
\author{Abhisek Bandyopadhyay$^{1,2}$}
\author{Atasi Chakraborty$^3$}
\author{Sayantika Bhowal$^{3,4}$}
\author{Vinod Kumar$^5$}
\author{M. Moretti Sala$^6$}
\author{A. Efimenko$^7$}
\author{F. Bert$^8$}
\author{P. K. Biswas$^9$}
\author{C. Meneghini$^{10}$}
\author{N. B\"{u}ttgen$^{11}$}
\author{I. Dasgupta$^3$}
\author{T. Saha Dasgupta$^{12}$}
\author{A. V. Mahajan$^5$}
\author{Sugata Ray$^{1}$}
\email{mssr@iacs.res.in}
\affiliation{$^1$School of Materials Sciences, Indian Association for the Cultivation of Science, 2A \& 2B Raja S. C. Mullick Road, Jadavpur, Kolkata 700 032, India}
\affiliation{$^2$Department of Physics, Indian Institute of Science Education and Research, Dr. Homi Bhabha Road, Pune 411 008, India}
\affiliation{$^3$School of Physical Sciences, Indian Association for the Cultivation of Science, 2A \& 2B Raja S. C. Mullick Road, Jadavpur, Kolkata 700 032, India}
\affiliation{$^4$Department of Physics and Astronomy, University of Missouri, Columbia, Missouri 65211, USA}
\affiliation{$^5$Department of Physics, Indian Institute Of Technology Bombay, Powai, Mumbai 400076, India}
\affiliation{$^6$Dipartimento di Fisica, Politecnico di Milano, P.zza Leonardo da Vinci 32, I-20133 Milano, Italy}
\affiliation{$^7$ESRF–The European Synchrotron, 71 Avenue des Martyrs, 38000 Grenoble, France}
\affiliation{$^8$Universit\'{e} Paris-Saclay, CNRS, Laboratoire de Physique des Solides, 91405, Orsay,France}
\affiliation{$^9$ISIS Facility, Rutherford Appleton Laboratory, Chilton, Didcot, Oxon OX110QX, United Kingdom}
\affiliation{$^{10}$Dipartimento di Scienze, Universit\'{a} Roma Tre, Via della Vasca Navale, 84 I-00146 Roma, Italy}
\affiliation{$^{11}$Experimental Physics V, Center for Electronic Correlations and Magnetism, University of Augsburg, 86159 Augsburg, Germany}
\affiliation{$^{12}$Department of Condensed Matter Physics and Material Sciences, S. N. Bose National Centre for Basic Sciences, Block JD, Sector 3, Saltlake, Kolkata -700106, India}
\begin{abstract}
Spin-orbit coupling (SOC) often gives rise to interesting electronic and magnetic phases in an otherwise ordinary pool of paramagnetic heavy metal oxides. In presence of strong SOC, assumed to be working in $j$-$j$ coupling regime, 5$d^4$ iridates are generally speculated to possess a nonmagnetic $J_{eff}$~=~0 singlet ground state, which invariably gets masked due to different solid-state effects (e.g. hopping). Here, we try to probe the trueness of the atomic SOC-based proposal in an apparently 1-dimensional system, Sr$_3$NaIrO$_6$, possessing a 2$H$ hexagonal structure with well separated Ir$^{5+}$ (5$d^4$) ions. But all the detailed experimental as well as theoretical characterizations reveal that the ground state of Sr$_3$NaIrO$_6$ is not nonmagnetic, rather accommodating a significantly high effective magnetic moment on Ir$^{5+}$ ion. However our combined dc susceptibility ($\chi$), ${}^{23}$Na nuclear magnetic resonance (NMR), muon-spin-relaxation/rotation ($\mu$SR) and heat capacity ($C_p$) measurements clearly refute any sign of spin-freezing or ordered magnetism among the Ir$^{5+}$ moments due to geometrical exchange frustration, while in-depth zero-field (ZF) and longitudinal field (LF) $\mu$SR investigations strongly point towards inhomogeneous quantum spin-orbital liquid (QSOL)-like ground state. In addition, the linear temperature dependence of both the NMR spin-lattice relaxation rate and the magnetic heat capacity at low temperatures suggest low-lying gapless spin excitations in the QSOL phase of this material. Finally, we conclude that the effective SOC realised in $d^4$ iridates are unlikely to offer a ground state which will be consistent with a purely $j$-$j$ coupling description.
\end{abstract}

\maketitle
Transition metal based oxide systems with heavier (4$d$ or 5$d$) elements possess strong spin orbit coupling (SOC) and crystal field splitting energy (${\Delta}$) with moderate Coulomb repulsion $U$, and have been at the centerstage of condensed matter physics research for more than a decade now. The reason for this is the growing excitement surrounding the realization of exotic electronic and magnetic ground states in such systems within these parameter ranges~\cite{Balents_review,SIO_PRL}, which have demanded a closer scrutiny. However, one big challenge has been to untie the complex, interconnected knot of several electronic parameters as every attempt of independently manipulating one crucial parameter invariably modulates others, and often in a direction adverse to the desired result.
Therefore, the everlasting search for a suitable material playground remains on, where favorable manipulations towards the desired properties can be carried out.

With this particular motive in mind, attempts are made here to probe a quasi-1 dimensional (1D) system, Sr$_3$NaIrO$_6$ (SNIO), consisting of significantly isolated Ir$^{5+}$ (5$d^4$) ions which may help to realize the $J_{eff}$~=~0 state within the purely atomic $jj$-coupling description and even the proposed van-Vleck type magnetic response~\cite{Khaliulin_PRL}. The structure of the compound, obtained from the Rietveld refinement of room temperature X-Ray Diffraction (XRD) pattern (Fig. S5 of Supplementary Materials(SM)), consistent with previous literature~\cite{ref_1}, is shown in Fig. 1(a). Clearly, this compound has a 1D-like structure in which alternating IrO$_6$ octahedra and trigonal prismatic NaO$_6$ units form a column(/chain) along the $c$-direction. The Ir-O-Na-O-Ir intrachain superexchange interactions are anticipated to be rather feeble, resulting a nearly isolated magnetic and electronic environment for Ir. The interchain Ir-Ir distances on the other hand are reasonably large (5.9~\AA~~and 6.78~\AA), and therefore, weak electronic and spin-spin interactions among them are expected. Such a situation naturally points towards narrower Ir bands, reduced hopping and consequently increased effective electronic correlations and SOC effects. This indeed boosts up the possibility of realizing the SOC dominated $J_{eff}$~=~0 state in SNIO.

But, it should be noted that such 1D systems aren't devoid of controversy regarding their true dimensionality~\cite{Lee_InorgChem_1999,sampat_prb_2005,sampat_prb_2011,SNRuO_prb_2016,calder_prb_2017,sampat_jpsj_2020}.
It has been shown recently~\cite{takegami_PRB_2020} that the $A_2B$IrO$_6$ iridates have essentially zero O~2$p$ to Ir~5$d$ charge-transfer energies leading to unusually strong covalent interaction and significantly effective Ir-O-O-Ir magnetic exchange interactions.
Most importantly, it has also been shown by Liu {\it et al.}~\cite{Liu_PRL_2012}, using resonant inelastic x-ray scattering (RIXS) experiments on Sr$_3$CuIrO$_6$, that the expectation of strong SOC limit in these systems could be deceptive.

Amid these controversies over true dimensionality and the actual strength of SOC in columnar iridates, our results in SNIO not only indicate breaking down of the possible $J_{eff}$~=~0 state, they also show that Ir$^{5+}$ possesses one of the highest magnetic moments ($\sim$ 0.5 $\mu_B$) in SNIO among the 5$d^4$ iridates. Our combined dc susceptibility, ${}^{23}$Na NMR and $\mu$SR studies, together with the {\it ab-initio} density functional theory (DFT) calculations confirm the true intrinsic nature of the developed large Ir moments in this compound. Despite having noticeable AFM interactions among these large local Ir$^{5+}$ moments ($\Theta_{CW}$ = -27 K), we show using detailed NMR, heat capacity and $\mu$SR characterizations that the SNIO fails to trigger any long-range magnetic order and even spin-glass freezing down to at least 40~mK (frustration parameter $\approx$ 675) due to enhanced quantum fluctuations as a result of geometrical exchange frustration arising from the triangular network (see Fig. 2(d)), and a gapless quantum spin-orbital liquid (QSOL) like dynamical magnetic ground state is stabilized. Lastly, considering the puzzling revelation of finite moment generation in this pseudo-1D $d^4$ iridate system Sr$_3$NaIrO$_6$, we go on to compare bandwidth {\it vs.} moment trends in the three columnar systems and also of a 6$H$ iridate Ba$_3$ZnIr$_2$O$_9$ (BZIO), which reveals presence of only a moderate SOC in all these systems where intermediate spin-orbit coupling limit appears to be valid.

\begin{figure}
\begin{center}
\resizebox{9cm}{!}
{\includegraphics[27pt,416pt][573pt,734pt]{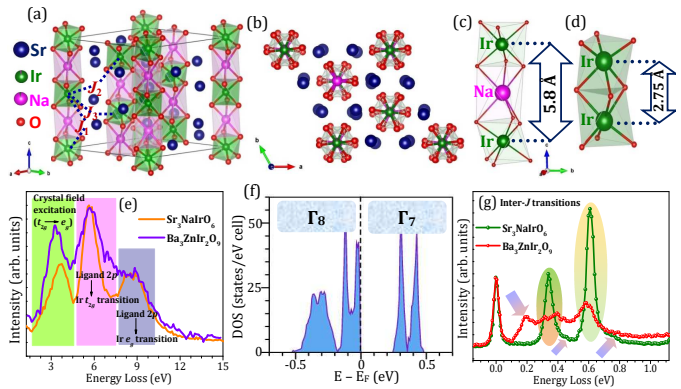}}
\caption{(colour online) (a) Crystal structure of SNIO with the intra- and inter-chain Ir-Ir exchange pathways (dotted blue lines); (b) Crystal structure, as viewed along $c$-axis, showing isolated arrangement of the 1-D chains; Ir-Ir hopping pathways for SNIO (c) and the 6$H$ Ba$_3$ZnIr$_2$O$_9$ (d) samples; (e) 300 K Low-resolution (LR) high energy Ir $L_3$-RIXS features for the SNIO and a 6$H$ hexagonal perovskite Ba$_3$ZnIr$_2$O$_9$; (f) The total density of states (TDOS) of SNIO within GGA$+$SOC scheme; (g) The High-resolution (HR) low-energy Ir $L_3$-RIXS spectra for SNIO, along with the 6$H$ Ba$_3$ZnIr$_2$O$_9$, showing a comparison in their excitations within SOC states.}
\end{center}
\end{figure}

We start by probing the true dimensionality of SNIO, the degree of IrO$_6$ octahedral distortion and the resultant noncubic crystal field, which will all be collectively manifested in the Ir-O bandwidth. Also, the extent of possible site-disorder among Na and Ir would be of importance because such a disorder can bring two Ir ions close to each other (Fig. S7), creating face-sharing Ir-connectivity and strong Ir-O-Ir superexchange interactions. This may indeed influence the magnetic response of the system,~\cite{Chen_prb_2017,Laguna-Marco_prb_2020} especially if the bulk magnetic response truly becomes nonexistent ($J_{eff}$~=~0 for Ir$^{5+}$ in perfectly atomic strong SOC limit). The pure 5+ valence of Ir in SNIO is confirmed by combined Ir 4$f$ core level x-ray photoelectron spectroscopy (XPS) and Ir $L_3$-x-ray absorption near edge structure (XANES) spectroscopic studies, which are resumed in Fig. S8.

Detailed XRD as well as Ir $L_3$-edge extended x-ray absorption fine structure (EXAFS) spectroscopy (Table-S1, Fig. S5, along with the Fig. S6, and Table-S2) firstly indicate about 8-10\% of Na/Ir antisite disorder (ASD) in the system. These experiments also reveal the distortion within the IrO$_6$ octahedra which has been summarized in Table-S1 (also see Fig. 1(b) and (c)). Such octahedral distortions are common in face-shared systems which brings forth further splitting to the transition metal $t_{\rm 2g}$ and $e_{\rm g}$ levels and modify the bandwidth.
Therefore next we compare the low resolution Ir-$L_3$ edge RIXS spectrum of SNIO (Fig. 1(e)) with the same from Ba$_3$ZnIr$_2$O$_9$ (BZIO)~\cite{our_prl_2016}, possessing face-shared isolated Ir$_2$O$_9$ dimers (Fig. 1(d)). The very first observation is the presence of much broader bands in the 6$H$ compound BZIO, which confirms that the effective spatial dimensionality is indeed substantially lesser in SNIO. The pseudo 1D nature of SNIO in the $a-b$~plane is shown in Fig. 1(b) while the enhanced Ir-Ir isolation along the $c$-direction in SNIO compared to BZIO is more clearly presented in Figs. 1(c) and (d). Given the fact that IrO$_6$ octahedra are more distorted in the 6$H$ system~\cite{our_prl_2016} which should lift the degeneracy of the $t_{\rm 2g}$ and $e_{\rm g}$ bands, it is no surprise that the peaks at 6~eV (O~2$p$~$\rightarrow$~Ir~$t_{\rm 2g}$) and 9~eV (O~2$p$~$\rightarrow$~Ir~$e_{\rm g}$) for the system are more wide and nearly overlapping. Due to the same reason the effective crystal field splitting energy also lowers down (peak around 3~eV) in the 6$H$ compound. The isolated nature of Ir$^{5+}$ ions in SNIO and comparatively lesser bandwidths offer a possibility to realize the nearly atomic SOC limit and for strong enough SOC,  $J_{eff}$ =0 ground state may be realised. Moreover, like the whole lot of iridates, this compound is also found to be insulating both by XPS valence band as well as electrical resistivity measurements (Fig. S9) which indicates clear prevalence of SOC in SNIO too.

In light of the above experimental observations, we have first carried out non-spin-polarized density functional theory (DFT) calculations using plane wave basis as implemented in the Vienna ab-initio simulation package (VASP)~\cite{vasp1,vasp2}(See SM for details). The consistency of the VASP calculations was checked using the N-th order muffin tin orbital method~\cite{NMTO} (See SM for details). Our calculation~\cite{NMTO} shows a $t_{2g}-e_{g}$ splitting of $\sim$3.5 eV  within the Ir-$d$ states, consistent with the RIXS measurements described earlier. The trigonal distortion of the IrO$_6$ octahedra, further, splits the $t_{2g}$ states into low lying $a_{1g}$ and higher lying doubly degenerate $e_g^{\pi}$ states with a relatively small non-cubic crystal field splitting of 0.05 eV. The nearly isolated Ir atoms in SNIO, lead to a significantly narrow Ir-$t_{2g}$ bandwidth of 0.7 eV as compared to other $d^4$ iridates, where the bandwidth is typically $1.2-1.4$ eV \cite{our_prl_2016,Nag_PRL,our_BYIO,Nag-6H-PRB}. Inclusion of SOC within the GGA+SOC \cite{vasp1,vasp2} formalism splits this narrow Ir-$t_{2g}$ states into completely filled single particle $j_{\rm eff}= 3/2$ ($\Gamma_8$) and empty $j_{\rm eff}={1/2}$ ($\Gamma_7$) states introducing an energy gap in the electronic structure [see Fig. 1 (f)] emphasizing the important role of SOC for the system to be insulating.

Next the low energy, high resolution Ir-$L_3$ edge RIXS spectra~\cite{Marco-rixs} from SNIO and BZIO are compared in Fig. 1(g). Clearly, there are substantial differences in spectral width, intensity, energy positions as well as number of spectral features (peak at 0.18~eV in BZIO and presence of weak asymmetry in the higher-energy side of the second and third peaks at $\sim$ 0.34 and 0.6~eV respectively in SNIO, shown by bicoloured arrows in Fig. 1(g)) between the two which could be associated with enhanced degree of hopping and greater extent of Ir $t_{\text{2}g}$ trigonal crystal field splitting in BZIO~\cite{Nag_PRL} compared to SNIO. Moreover, there would be bimodal distributions of local noncubic crystal fields, Ir-O covalent interactions, Ir-Ir superexchange interactions $J_{SE}$($\approx$ 4$t^2$/$U$) and the subsequent effective SOC strengths in SNIO due to the presence of $\sim$90\% Ir in the ordered sites along with the $\sim$10\% Ir in the antisite positions, which is naturally absent in BZIO. However, as any attempt to directly estimate effective SOC ($\lambda_{eff}$) from the low-energy Ir-$L_3$ RIXS data, assuming only atomic effects, is going to be futile~\cite{Nag_PRL,Revelli_SciAdv}, we refrain from making such an estimate here. Nonetheless, similarity between the low-energy RIXS spectra of SNIO and BYIO~\cite{our_BYIO} confronts that the IrO$_6$ octahedra are significantly isolated in SNIO too, even though the interaction pathways are substantially different.

Next comes the crucial question of whether the elusive $J_{eff}$~=~0 state is realized in SNIO, or not. Bulk susceptibility ($\chi$) {\it vs.} temperature ($T$) study (Fig. 2(a)) and the consequent Curie-Weiss fit ($\chi$($T$) = $\chi_0$ + $\frac{C}{T-\Theta_{CW}}$) give the most direct answer to this question. In contrary to the ideal $J_{eff}$ = 0 situation we observe a high effective moment $\mu_{eff}$ $\approx$ 0.5~$\mu_B$/Ir (where, Curie constant $C$ $\approx$ 0.031 K) in SNIO with a Weiss temperature $\Theta_{CW}$ of -27~K. This observation directly refutes a single ion description of Iridium and endorses the previous results on the isostructural compounds~\cite{Lee_InorgChem_1999,sampat_prb_2011,SNRuO_prb_2016,sampat_jpsj_2020}, depicting SNIO as a 3-dimensional (3D) system. Moreover, the large value of magnetic moment raises a doubt about the high SOC, $jj$-coupling ground state picture, because in this case the prediction of only intermediate $\lambda_{eff}$ in columnar systems~\cite{calder_prb_2017,Liu_PRL_2012} seem to be more suitable. Additionally, it should also be noted that the existing Na/Ir site-disorder may also affect the magnetic response to a significant level, as mentioned earlier. Interestingly, spin polarized calculations with the spin quantization axis along [001] direction shows that an AFM configuration with the anti-parallel spin orientations at the neighboring Ir sites within each chain is slightly lower in energy ($\Delta E$/f.u. =3 meV) compared to the non-magnetic structure. This indicates the presence of AFM interactions in SNIO, consistent with the negative Weiss temperature extracted from the Curie-Weiss fit of the susceptibility measurements. The spin and orbital moments at the Ir site in the computed AFM structure are respectively 0.62 $\mu_B$ and 0.26$\mu_B$ in contrast to a non-magnetic $J_{eff}$=0 state, as proposed in a recent theoretical study by  Ming {\it et. al}~\cite{snio_prb}.
\begin{figure}
\begin{center}
\resizebox{8.6cm}{!}
{\includegraphics[33pt,336pt][548pt,760pt]{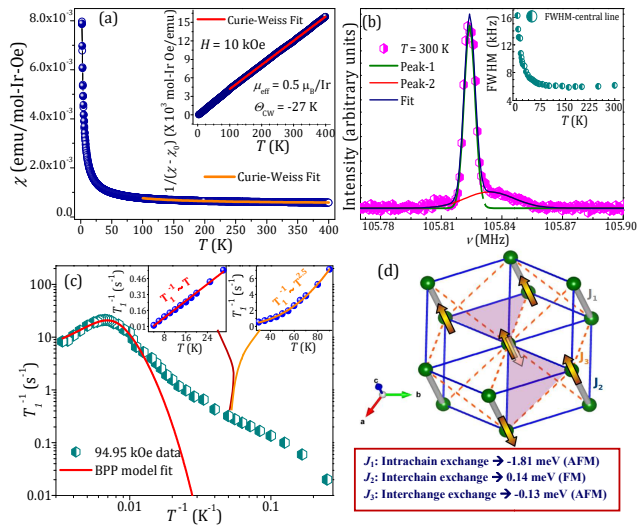}}
\caption{(colour online) (a) ZFC(open circles) and FC(shaded circles) dc magnetic susceptibilities at 10 kOe magnetic field along with the Curie-Weiss fitting; (b) The 300 K ${}^{23}$Na NMR spectrum along with the fit, Inset: $T$-variation of the central NMR linewidth (c) $T$-evolution of the NMR spin-lattice relaxation rate, Inset: power law dependency of the same below 90~K, (d) The edge-shared triangular network (light pink shaded region) formed by Ir ions of the neighboring chains;}
\end{center}
\end{figure}

The other important issue in such weakly magnetic materials, could be that the bulk magnetization may also be influenced by the extrinsic factors such as small paramagnetic impurities, which often lead to the low temperature Curie-Weiss-like upturn on top of the intrinsic magnetic signal~\cite{byionmrprb}. So we employed Nuclear Magnetic Resonance (NMR) as a complementary tool for probing intrinsic magnetism in this system. Consequently,  ${}^{23}$Na ($I$ = 3/2) NMR has been performed in a fixed field of 94.95 kOe and in the temperature range of 4-300 K and the results have been summarized in Figs. 2(b)-(c) and S10(a)-(c). At room temperature, there is a clear knee on the high-frequency side [Fig. 2(b)] of the resonance lineshape, where a fit with a combination of two Gaussians is found to be satisfactory.  The relative intensity of the Gaussian centred on the knee is about 30 \% while the mainline constitutes about 70 \% of the spectral intensity.  The peak positions and the relative intensities are found to be nearly temperature independent. There is a gradual broadening of the lines with decrease in temperature and below about 30 K, the knee is no more distinctly seen~\cite{MISC1}. The presence of the second NMR signal (`knee' feature) is attributed to Na/Ir site-disorder. In the ideal structure, the Na and Ir ions alternate to form a chain.  Each Ir (Na) has then two Na (Ir) neighbors. In the case of site disorder ($x \%$) in the dilute limit, one will have a probability of finding a Na with no Ir neighbours to be $x^2$ while that with only one Ir neighbor will be $2x(1-x)$, and the remaining fraction $(1-x)^2$ will be unaffected.  In the dilute limit, then, a $2x\%$ intensity will appear in the second resonance which is likely to be shifted due to a different chemical and magnetic environment. For higher $x$, other chemical environments become possible and the mainline intensity gets depleted further.  Given that a 10 \% site disorder cannot be ruled out from our structural analysis (both XRD and EXAFS), a 30 \% intensity appearing in the `knee' feature is consistent with Na/Ir site disorder. The variation of the $^{23}$Na NMR-shift and linewidth from the two-Gaussian fits between 30 K and 300 K are shown in Figs. S10(a) and (b). The near absence of any variation in the NMR-shift with temperature suggests weak hyperfine coupling of the ${}^{23}$Na with magnetic Ir, or that the moment on Ir is weak, or perhaps both. But, an increase in central NMR mainline-width at low temperatures (see inset to Fig. 2(b)) scales with the low-temperature upturn of the $\chi$($T$) data (Fig. 2(a)). This affirms intrinsic nature of the observed large local Ir$^{5+}$ magnetic moments in SNIO, as such increase in linewidth arises from a varying dipolar coupling interactions to the Ir$^{5+}$-local moments. Therefore, almost temperature-independent NMR-shift could certainly be attributed to the weak hyperfine coupling of ${}^{23}$Na with Ir$^{5+}$ moments due to Ir 5$d$-O 2$p$ orbital overlap.

In order to understand the magnetic interactions between the large intrinsic Ir spins, we have carried out DFT calculations. For a quantitative estimation of the Ir-Ir exchange couplings, we have calculated the symmetric exchange interactions $J^s$ by mapping the total energies, computed within GGA+SOC+U, of several spin configurations to the Heisenberg spin model $H_{spin} = \sum_{ij} J^{s}_{ij} (\vec{S_i} \cdot \vec{S_j})$. Our calculations show that the dominant interaction $J_1$ corresponding to the nearest neighbor (nn) intra-chain Ir-Ir coupling is AFM in nature with a magnitude of 1.8 meV. The second and third nn interactions, $J_2$ and $J_3$, corresponding to the inter-chain Ir-Ir interactions, are on the other hand weaker, and are respectively FM and AFM in nature. Interestingly, both $J_2$ and $J_3$ have six nn and their magnitudes are also comparable, viz., $J_2$= 0.14 meV and $J_3$= 0.13 meV, resulting in a frustrated triangular network as indicated in Fig. 2(d), prohibiting any long range magnetic order in the system.

To experimentally probe the local internal magnetic fields and spin dynamics of this sample, we have performed $\mu$SR experiments. The measured time evolution of the muon polarization $P$($t$)~\cite{MISC2} in zero external field is shown in Fig. 3(a) for some selected temperatures from 40~mK to 300~K. Remarkably, upon cooling down, a fast initial relaxation develops as can be seen by the loss of initial polarisation. Meanwhile, the longer time relaxation remains moderate and hardly changes throughout the whole temperature range. To monitor this evolution we could fit the ZF polarization curves at all temperature with the minimal model:
\begin{equation}
P(t) = A_1\exp(-\lambda_1 t) + (1-A_1)\exp(-\lambda_2 t)
\label{Eq_musr}
\end{equation}
where the first exponential component accounts for the fast front end with a relative weight $A_1$ and the second component stands for the slower relaxing tail. The relaxation rate $\lambda_2$ for the latter slow component is small and weakly $T$ dependent, varying between 0.31(5) to 0.42(1)$\mu$s$^{-1}$ (see Fig. S10(d)) in the whole temperature range. The temperature evolution of the relaxation rate $\lambda_1$ and the weight of this fast relaxing component are presented on Fig. 3(b). Below about 100~K, there is a sharp increase of $\lambda_1$ demonstrating the slowing down of the spin-fluctuations, likely due to the enhancement of spin-correlations, for an increasing volume fraction $A_1$ of the sample. At still lower temperatures, the spin dynamics  strikingly reaches a $T$ independent regime from about 30~K down to the lowest measurement at 0.04 K. Despite such obvious slowing down of the spin dynamics, the system does not undergo any magnetic transition and the spins continue to fluctuate far below the magnetic interaction energy scale $k_B \theta$. The absence of `1/3rd tail' and of spontaneous oscillations down to  40~mK, confirms the dynamic nature of the local fields in the ground state of SNIO. Moreover, since the fraction $A_1$ evolves with $T$, the two components in the model do not stand for 2 muon sites. Rather, in the low $T$ regime, $A_1 \sim 0.5$  points at a strong inhomogeneity of the ground state.

In order to get further insight into the origin of the relaxation observed at low temperatures, we have investigated the evolution of the muon polarization at $T$ = 0.08~K under several external longitudinal fields, as shown in Fig. S11. Besides the fast decoupling for applied fields $<20$~G of a static component from nuclear fields, the recovery of the polarization appears very gradual with sizeable decoupling for low fields but also persisting relaxation up to 3000~G, in line with a dynamical relaxation. The gradual decoupling could not be reproduced with a simple two dynamical components model. This suggests again a broad distribution of internal fields and fluctuations in the ground state for which Eq.~\ref{Eq_musr} is a too simplistic approximation.
\begin{figure}
\begin{center}
\resizebox{8.6cm}{!}
{\includegraphics[20pt,338pt][560pt,761pt]{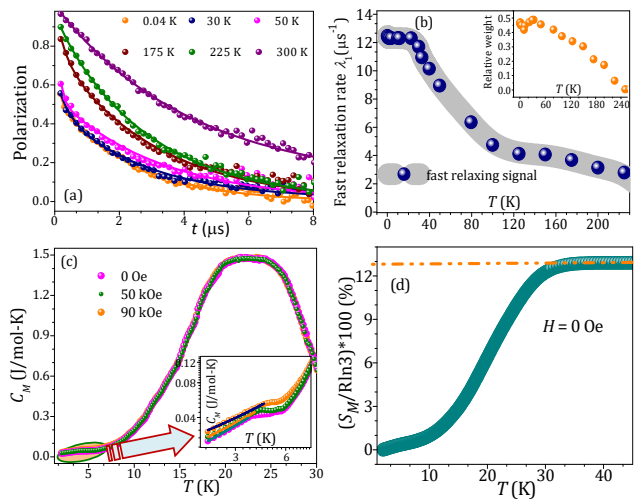}}
\caption{(colour online) (a) Time evolution of ZF muon polarization; (b) Temperature dependence of the fast relaxation rate; Inset: Temperature variation of the corresponding weight fraction ; (c) Temperature dependence of magnetic specific heat $C_M$ at zero field and applied magnetic fields, Inset: expanded views of the linear $C_M$ versus $T$ variations at low-$T$; (d) Zero field magnetic entropy $S_M$}
\end{center}
\end{figure}

Indeed, as a result of the near about 8-10\% Na/Ir anti-site disorder (as obtained from structural analysis), there would be a distribution of local iridium magnetic moments, which with their individual local geometries within the structure, support the strongly inhomogeneous nature of QSL ground state in this material.

The persisting spin-fluctuations should also be reflected in the magnetic entropy. To further explore the magnetic ground state, the heat capacity measurements ($C_p$ versus $T$) have been performed in zero field and several applied magnetic fields (Fig. S12(a)). No sharp $\lambda$-like anomaly, hallmark of thermodynamic phase transition, was observed in the measured temperature range of $C_p$. However below about 10 K, presence of a weak `hump'-like feature and its magnetic field-dependence suggest a two-level Schottky anomaly, possibly arising from the uncompensated site-disordered Ir$^{5+}$ magnetic centers~\cite{ba3irti2o9,syioprb,btrhirprb}. So, after subtracting the lattice ($C_{lattice}$) and the two-level Schottky ($C_{sch}$) contributions (detailed analysis are resumed in SM), the $T$-variations of the magnetic specific heat $C_M$ are plotted in Fig. 3(c), clearly pointing towards a weak magnetic field-dependence of $C_M$ in the 2$-$10~K range. Such field-dependence supports short-range magnetic correlations in this system. Further, $C_M$ displays a broad maximum in the 20-25 K $T$-region, which remains unaltered under applied magnetic fields, supporting highly frustrated nature of SNIO like many other reported spin-liquids~\cite{5000nag18,li2rho3prb57,khuntia-prb-Li2RhO3}. In addition, the $C_M$ has a finite $T$-linear contribution ($C_M$ $\sim$ $\gamma$$T$ with $\gamma$ $\approx$ 10 mJ/mol-K$^2$, shown in the inset of Fig. 3(c)) at very low-$T$, unusual for charge insulators. This implies low-energy gapless spin excitations or the presence of metal-like spinon Fermi surface in SNIO, as discussed in gapless quantum spin liquid candidates~\cite{our_BYIO,5000nag18,khuntia-prb-Li2RhO3,nagprbdp29,nagprbdp41,nagprbdp44,cava-Ba4NbIr3O12-prm23,cava-Ba4NbIr3O12-prm24,H3LiIr2O6-nature,cava-Ba4NbIr3O12-prm}. Similar gapless nature of spin excitations of SNIO is also corroborated from the power-law dependency of the ${}^{23}$Na NMR spin-lattice relaxation rate ($T_1^{-1}$ versus $T$) below 90 K (inset to Fig. 2(c)), which can be considered as a fingerprint of gapless QSL material~\cite{khuntia-prb-Li2RhO3,Ba3InIr2O9-PRB2017,khuntia-prb-Li2RhO3-18}. Finally the release of magnetic entropy $S_M$ (Fig. 3(d)) has been found to be only $\sim$ 12.6\% of the maximum $R$$\ln$3 ($\approx$ 9.134 J/mol-K), affirming persistence of spin fluctuation and also low-energy spin excitations in this compound~\cite{cava-Ba4NbIr3O12-prm23,cava-Ba4NbIr3O12-prm24,H3LiIr2O6-nature,cava-Ba4NbIr3O12-prm}.

Overall, the magnetic entropy starts to decrease from around 30 K with temperature lowering (Fig. 3(d)), which is exactly the temperature below which there is complete absence of $T$-evolution of the $\mu$SR relaxation (Fig. 3(b)), and also close to the $T$-region where the magnetic field independent broad maximum appears in the $C_M$ data (Fig. 3(c)), and linear $T$-dependent NMR spin-lattice relaxation rate (inset to Fig. 2(c)) becomes evident, which point towards the emergence of rather unconventional inhomogeneous low-$T$ ($\sim$ $<$ 30K) spin-liquid phase in SNIO.

\begin{figure}
\begin{center}
\resizebox{8.6cm}{!}
{\includegraphics[179pt,406pt][356pt,594pt]{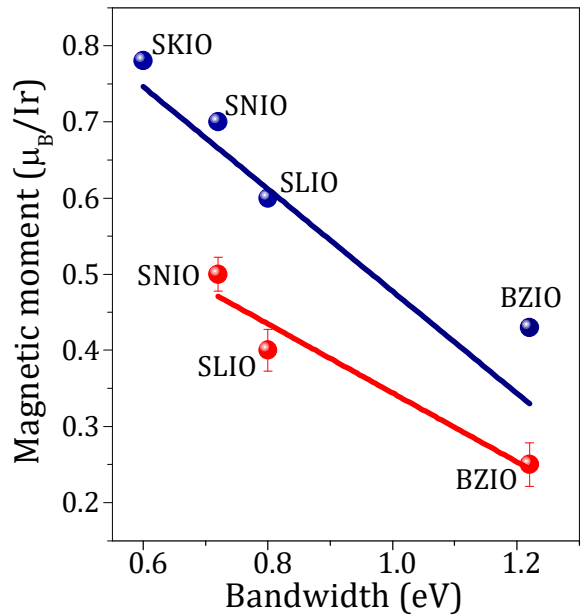}}
\caption{(colour online) Bandwidth dependence of effective Ir-magnetic moments, obtained from experiments (red shaded circles with error bars) and calculations (blue shaded circles), of three columnar compounds (SNIO, SLIO: Sr$_3$LiIrO$_6$ and SKIO: Sr$_3$KIrO$_6$) and a 6$H$ Ba$_3$ZnIr$_2$O$_9$.}
\end{center}
\end{figure}
Finally, we show the dependence of the calculated (blue circles) and experimentally observed (red circles, obtained from the Curie constants of dc susceptibility data) magnetic moments as a function of theoretically predicted bandwidth from three columnar systems and BZIO in Fig. 4 where it is clearly seen that the magnetic moment continues to decrease when the bandwidth gradually increases. The enhancement of magnetic moment with reduction of bandwidth can only be explained provided the spin-orbit coupling is weak. In particular in a solid the strength of the SOC is renormalised as $\lambda_{eff}$ = $\lambda_{atomic}/S$, where $S$ is the total spin. For $d^4$ systems with $S$=1, SOC is much weaker in comparison to $d^5$ systems with $S$=1/2, providing further credence that SOC for $d^4$ systems is always in the intermediate coupling regime.

Overall, our detailed experimental and theoretical results on a pseudo 1D columnar Iridate, Sr$_3$NaIrO$_6$ with Ir$^{5+}$ oxidation state, reveal that this system fails to realize a $J_{eff}$~=~0 state, resulting in a rather large local moment on Ir due to SOC in the intermediate coupling regime, despite the significant isolation of Ir ions within the structure. Interestingly, the geometrical frustration also turns out to be extremely strong due to the comparable strengths of nearest- and next-nearest-neighbor Ir-O-O-Ir interchain exchange couplings and as a result the system doesn't undergo any magnetic transition down to at least 40~mK, even though the short-range spin-spin correlation starts to dominate from a temperature as high as 30~K.

A.B. thanks CSIR, India and IACS for fellowship. A.B. also acknowledges SERB, DST, INDIA for National Postdoctoral Fellowship (N-PDF, File NO. PDF/2020/000785). S.R. thanks Technical Research Center of IACS, Indo-Italian POC for support to carry out experiments in Elettra, Italy, and Jawaharlal Nehru Centre for Advanced Scientific Research and SINP from DST-Synchrotron-Neutron project for performing experiments at ESRF (Proposal No. HC-2872) and ISIS-RAL, UK. S.R. also thanks SERB, DST for funding (Project No. CRG/2019/003522). I.D. thanks Science and Engineering Research Board (SERB), India
(Project No.EMR/2016/005925) and Department of Science and Technology-Technical Research Centre (DST-TRC) for financial support. TS-D acknowledges J.C.Bose National Fellowship (grant no. JCB/2020/000004) for funding.

\newpage

\section{Supplementary material(SM)}

\subsection{Methodology}
\subsubsection{Experimental details}
Polycrystalline Sr$_3$NaIrO$_6$ sample was synthesized using conventional solid state reaction technique using stoichiometric amounts of high purity ($>$ 99.9\%) predried SrCO$_3$, Na$_2$CO$_3$ and IrO$_2$ as starting materials. These starting materials were homogeneously ground, followed by initially calcination at 700$^{\circ}$ in air for 12 hours, and finally the resultant mixture was reground and annealed at several higher temperatures (800$^{\circ}$ and 900$^{\circ}$ for 12 hours each) in air with few intermediate grindings. The phase purity of the sample was checked from x-ray powder diffraction measured at Bruker AXS: D8 Advance x-ray diffractometer with Cu K$_{\alpha}$ radiation at the room temperature. The obtained XRD data was analyzed by using the Rietveld technique and the structural refinement was performed by FULLPROF program~\cite{fullprof}. The Ir $L_3$-edge x-ray absorption near-edge structure (XANES) and extended x-ray absorption fine structure (EXAFS) measurements of this sample have been performed at the XAFS beamline of Elettra (Trieste, Italy) synchrotron radiation facility at room temperature in standard transmission geometry. Data treatment and quantitative analysis of EXAFS have been performed using the freely available Demeter package~\cite{artemis,ravel} (Athena \& Arthemis programs) using atomic clusters generated from crystallographic structure. To verify chemical homogeneity and any cation-off-stoichiometry in the sample, energy dispersive x-ray (EDX) analysis was also performed using a field emission scanning electron microscope (JEOL, JSM-7500F). The cation-stoichiometry was also checked by inductively coupled plasma-optical emission spectroscopy (ICP-OES) using a Perkin Elmer Optima 2100 DV instrument. The x-ray photoemission spectroscopy (XPS) measurements were carried out using an Omicron electron spectrometer, equipped with Scienta Omicron sphera analyzer and Al K$_{\alpha}$ monochromatic source with an energy resolution of 0.5 eV. Before the spectra collection, the sample surface cleaning was achieved by optimized sputtering with argon ion bombardment to remove any kind of surface oxidization effect and the presence of environmental carbons in the pelletized samples. The collected spectra were processed and analyzed with the KOLXPD program. Electrical resistivity was measured by standard four-probe method within temperature range of 150 $-$ 400 K in a physical property measurement system (PPMS, Cryogenic). Magnetization measurements for this sample were carried out in the temperature range 2 $-$ 300 K and in magnetic fields up to $\pm$50 kOe in a superconducting quantum interference device (SQUID) magnetometer, Quantum Design. Further, heat capacity was measured within 2 $-$ 300 K temperature range in a physical property measurement system (PPMS, Quantum Design) in the zero field as well as several applied higher magnetic fields (upto 90 kOe). The muon-spin-rotation/relaxation ($\mu$SR) experiments, as local magnetic probe, were conducted with the EMU spectrometer at the ISIS large scale facility both in a helium flow cryostat and a dilution fridge. Moreover, the Ir $L_3$ -edge resonant inelastic x-ray scattering (RIXS)~\cite{Marco-rixs-sm} has been measured for this sample at the ID20 beamline of the European Synchrotron Radiation Facility (ESRF) using $\pi$-polarized photons and a scattering geometry with 2$\theta$ $\backsimeq$ 90$^{\circ}$ to suppress elastic scattering.
\subsubsection{Theoretical details}
The electronic structure calculations based on density functional theory (DFT) presented in this paper are carried out in the plane-wave basis within generalized gradient approximation (GGA)\cite{gga-sm} of the Perdew-Burke-Ernzerhof exchange correlation supplemented with Hubbard U as encoded in the Vienna \textit{ab-initio} simulation package (VASP) \cite{vasp1-sm,vasp2-sm} with projector augmented wave potentials \cite{aug1-sm,aug2-sm}. The calculations are done with usual value of U and Hund’s coupling (J$_H$) chosen for Ir with U$_{eff}$($\equiv$U-J$_H$) = 1.5 eV in the Dudarev scheme \cite{dudarev-sm}. In order to achieve convergence of energy eigenvalues, the kinetic energy cut off of the plane wave basis was chosen to be 550 eV. The Brillouin-Zone integrations are performed with 8$\times$8$\times$8 Monkhorst grid of k-points. The symmetry protected ionic relaxation of the experimentally obtained crystal structure has been carried out within VASP calculation using the conjugate-gradient algorithm until the Hellman-Feynman forces on each atom were less than the tolerance value of 0.01 eV/\AA.
%
%
\par The t$_{2g}$ crystal field splitting of our low-energy Hamiltonian, are obtained retaining Ir-t$_{2g}$ within the basis set and downfolding higher degrees for freedom, using the muffin-tin orbital(MTO) based Nth order MTO (NMTO) method \cite{nmto1-sm,nmto2-sm,nmto3-sm} as implemented in Stuttgart code. The NMTO method, relies on the self-consistent potentials borrowed from the linear MTO (LMTO) calculations \cite{lmto-sm}. For the self-consistent LMTO calculations within the atomic sphere approximation (ASA), the space filling in the ASA is obtained by inserting appropriate empty spheres in the interstitial regions.

\subsection{Structural Characterization and Composition Verification}
The collected room temperature XRD pattern (Fig. S5) has been satisfactorily refined with pure single phase having rhombohedral space group, $R\bar{3}c$. The refined crystal structure clearly demonstrates infinite chains of alternating NaO$_6$ trigonal prisms and IrO$_6$ octahedra along the $c$-direction. These chains are well separated from each other by nonmagnetic Sr cations. Further, the structural refinement clearly rules out any possibility of site-mixing between Sr (18$e$) and Na (6$a$), in line with the previously reported crystal structure~\cite{sniojmc-sm} of this compound.
\renewcommand{\thefigure}{S\arabic{figure}}
\begin{figure}
\begin{center}
\resizebox{8cm}{!}
{\includegraphics[131pt,257pt][422pt,488pt]{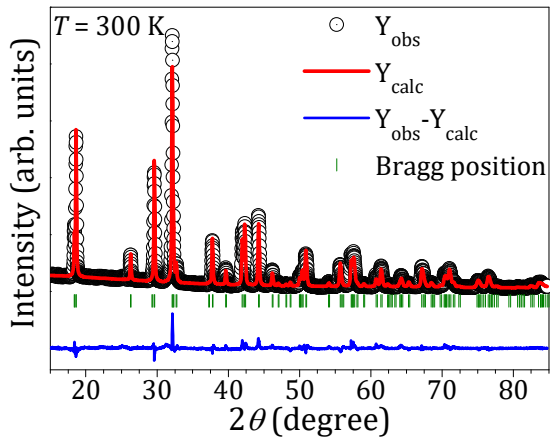}}
\caption{Rietveld refined XRD pattern of Sr$_3$NaIrO$_6$.}
\end{center}
\end{figure}
\renewcommand{\thetable}{S\arabic{table}}
\begin{table}[b]
\begin{center}
\caption{Structural information and agreement parameters obtained from Rietveld refinement of XRD data of Sr$_3$NaIrO$_6$ at 300 K. Space group: $R\bar{3}$c, $a$ = $b$ = 9.6487(5) {\AA}, $c$ = 11.5953(2) {\AA}, $\gamma$ = 120$^{\circ}$, $V$ = 934.8797 {\AA}$^3$, $R_p$ = 16.5, $R_{wp}$ = 15.5, $R_{exp}$ = 6.39, $\chi^2$ = 5.85, $\angle$O-Ir-O = 91.68$^{\circ}$ (obtuse) while $\angle$O-Ir-O = 88.32$^{\circ}$ (acute)\\}
\resizebox{8.6cm}{!}{
\begin{tabular}{| c | c | c | c | c | c | c |}
\hline Atoms & site & occupancy &  $x$ & $y$ & $z$ & $B$ {\AA}$^2$ \\\hline
  Sr & 18e & 1.0 & 0.3558(4) & 0 & 0.25 & 0.007(3)\\
  Na1 & 6a & 0.92-0.9 & 0 & 0 & 0.25 & 0.003(2)\\
  Ir1 & 6a & 0.08-0.1 & 0 & 0 & 0.25 & 0.003(2)\\
  Ir2 & 6b & 0.92-0.9 & 0 & 0 & 0 & 0.002(5)\\
  Na2 & 6b & 0.08-0.1 & 0 & 0 & 0 & 0.002(5)\\
  O & 36f & 1.0 & 0.1797(8) & 0.02298(6) & 0.1041(2) & 0.08(2)\\
\hline
\end{tabular}}
\end{center}
\end{table}
On top of it, a maximum of 8-10\% Na/Ir antisite-disorder (Table-S1) clearly reveals the consequent development of face-sharing Ir-Ir connectivity (Fig. S7(b)). The rotational distortion of the IrO$_6$ octahedra, in terms of the deviation of O-Ir-O bond angles (91.68$^{\circ}$) from the perfectly cubic 90$^{\circ}$ case, causes trigonal splitting in the triply degenerate Ir-$t_{\text{2}g}$ orbitals of this compound.
\renewcommand{\thefigure}{S\arabic{figure}}
\begin{figure}
\begin{center}
\resizebox{8cm}{!}
{\includegraphics[10pt,66pt][582pt,757pt]{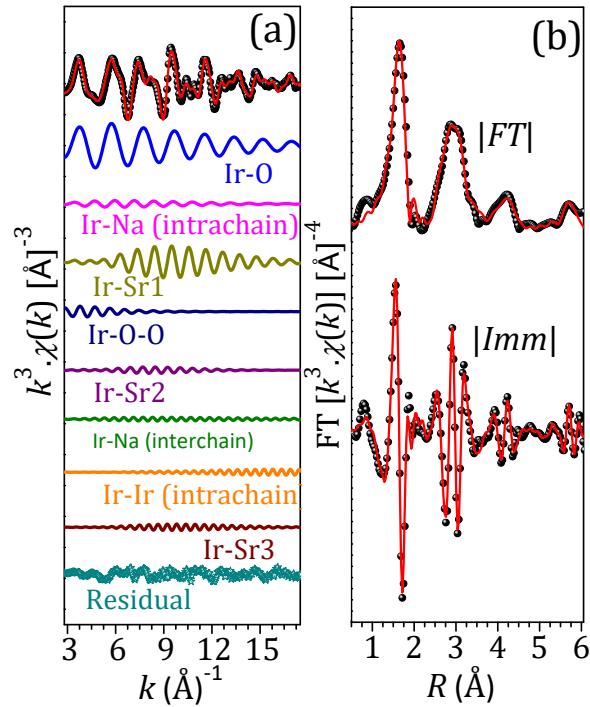}}
\caption{ (Left Panel) Ir $L_3$-edge $k^3$ weighted experimental EXAFS data (shaded black circles) and the respective best fit (red solid line) (a). The contributions from the individual single and multiple scattering paths (solid colored line) and the residual [$k^3$$\chi_{exp}$ $-$ $k^3$$\chi_{th}$] (open cyan stars) are also indicated, vertically shifted for clarity. (Right Panel) Fourier Transform of the respective experimental (shaded black circles) and theoretical (solid red line) curves (b); the magnitude ($|FT|$) and the imaginary parts (Imm) are illustrated; vertically shifted for clarity.}
\end{center}
\end{figure}

\renewcommand{\thetable}{S\arabic{table}}
\begin{table}
\begin{center}
\caption{The structural parameters obtained from the Multi-shell analysis of the Ir $L_3$-edge EXAFS spectra. The absolute mismatch between the experimental data and the best fit is $R$$^{2}$ = 0.024.\\ \\}
\resizebox{8.6cm}{!}{
\begin{tabular}{| c | c | c | c | c |}
\hline Shell & $N$ & $\sigma$$^{2}$ ($\times$ 10$^{2}${\AA}$^2$) & $R$({\AA}) & $R$$_{XRD}$({\AA}) \\\hline
   Ir-O & 6.0$^{\ast}$ & 0.30(1) & 1.97(2) & 2.03\\
   Ir-Na/Ir & 1.8-1.84/0.2-0.16 & 0.48(3) & 2.92(6) & 2.9\\
   (intrachain) &  &  &  &  \\
   Ir-Sr1 & 6.0$^{\ast}$ & 0.61(4) & 3.24(1) & 3.26\\
   Ir-O-O (MS) & 6.0$^{\ast}$ & 0.52(2) & 4.01(4) & 4.21\\
   Ir-Sr2 & 6.0$^{\ast}$ & 1.15(8) & 4.35(2) & 4.49\\
   Ir-Na & 6.0$^{\ast}$ & 0.26 & 5.60(7) & 5.65\\
   (interchain) &  &  &  &  \\
   Ir-Ir & 2.0$^{\ast}$ & 0.26(1) & 5.81(5) & 5.79\\
   (intrachain) &  &  &  &  \\
   Ir-Sr3 & 6.0$\ast$ & 0.64(7) & 5.72(8) & 5.75\\
\hline
\end{tabular}}
\end{center}
\end{table}

In order to further verify the local atomic distributions along the 1-D chain of this compound the local structure has been probed by performing Ir $L_3$-edge EXAFS measurement. The experimental EXAFS data, along with the respective fitting are shown in Fig. S6(a) and (b). The results obtained from the EXAFS data analysis are summarized in Table-S2 which supports the hypothesis of Na/Ir anti-site-disorder ($\sim$ 8-10\%) in this sample, in agreement with the XRD refinement.

The stoichiometry has been checked with SEM-EDX and it confirmed chemical homogeneity of our sample. In addition, the cation stoichiometry has been identified to be retained at the target composition, {\it i.e.,} Sr : Na : Ir being very close to 3 : 1 : 1 within the given accuracy of the measurement. The cation-stoichiometry was further quantified through ICP-OES analysis which also shows the actual stoichiometry to be almost close to the desired composition.

\renewcommand{\thefigure}{S\arabic{figure}}
\begin{figure}
\centering
\resizebox{8.6cm}{!}
{\includegraphics[20pt,540pt][545pt,734pt]{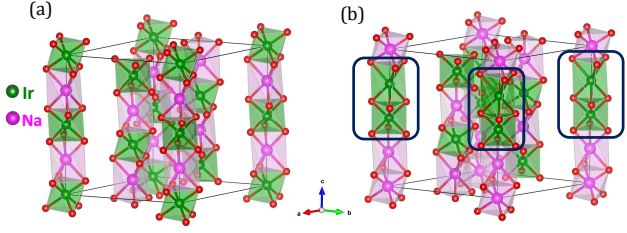}}
\caption{Structure of Sr$_3$NaIrO$_6$ with (a) fully ordered NaO$_6$ trigonal prisms (violet site) and IrO$_6$ octahedra (green site) along $c$-axis, and (b) some antisite disordered Na/Ir regions, caused by site-exchange between Na (0,0,$\frac{1}{4}$) and Ir (0,0,0). The Ir-Ir face sharing connectivity are shown by blue rectangular boxes. The bigger sized Sr cations are removed for the sake of clarity in both the figures.}
\end{figure}

\renewcommand{\thefigure}{S\arabic{figure}}
\begin{figure}
\resizebox{8.6cm}{!}
{\includegraphics[17pt,461pt][571pt,703pt]{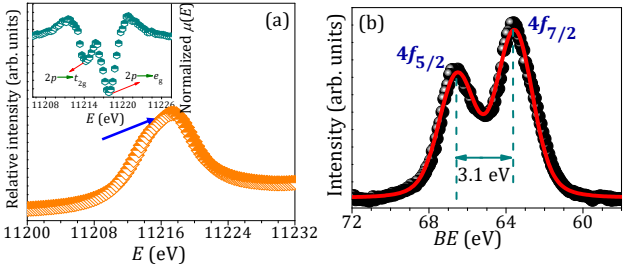}}
\renewcommand{\thefigure}{S\arabic{figure}}
\caption{(a) Ir $L_3$-edge XANES spectrum (half-filled orange triangle); Inset: respective second derivative curve (half filled cyan hexagon); (b) Ir 4$f$ core level XPS spectra (shaded black circles) along with the fitting (red solid line).}
\end{figure}
\subsection{Electronic characterization}
In order to confirm the Ir-valence in this sample, the measured Ir $L_3$-XANES spectrum is shown in Fig. S8(a). The asymmetric structure of the spectral line, in the form of a weak shoulder in the lower energy side (shown by solid blue line in Fig. S8(a)), resembles quite well with the Ir$^{5+}$ oxidation state~\cite{irxanes1-sm,irxanes2-sm}. Further, the corresponding second derivative curve (inset to Fig. S8(a)) clearly indicates the presence of 2$p$ $\rightarrow$ $t_{\text{2}g}$ (lower energy feature) and 2$p$ $\rightarrow$ $e_g$ (higher energy peak) transitions. The peak shape, structure and the peak intensity corresponding to 2$p$ $\rightarrow$ $t_{\text{2}g}$ transition affirm the pentavalent Ir~\cite{irxanes1-sm,irxanes2-sm} in this compound.

In addition, the energy positions of the 4$f_{7/2}$ and 4$f_{5/2}$ features in the measured Ir 4$f$ core level XPS spectrum [Fig. S8(b)] along with their spin-orbit separation of 3.1 eV, confirm pure 5+ charge state of Ir~\cite{nagprl-sm,nagprbdp-sm} in this sample. The presence of pure Ir$^{5+}$ species therefore serves as an indirect proof to refute any kind of noticeable oxygen defects in the SNIO sample, as change in oxygen-stoichiometry would eventually affect the Ir-oxidation state, and thereby, the spectral nature of both the Ir-$L_3$ XANES and Ir 4$f$ core-level XPS.

\begin{figure}
\resizebox{8cm}{!}
{\includegraphics[92pt,410pt][428pt,647pt]{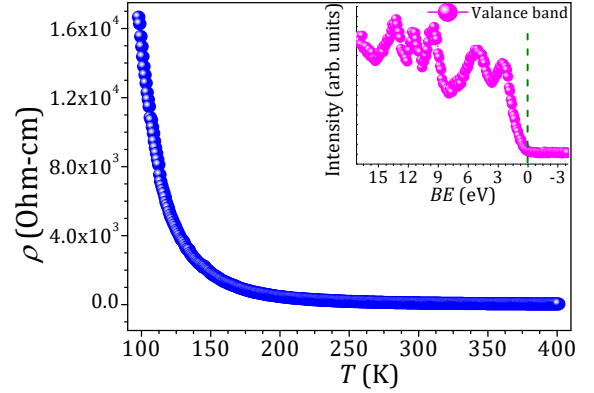}}
\renewcommand{\thefigure}{S\arabic{figure}}
\caption{Temperature dependent electrical resistivity variation (shaded blue circles); Inset: XPS valance band spectra (shaded violet circles); vertical green dashed line represents the Fermi energy position $E_F$.}
\end{figure}
\subsection{${}^{23}$Na nuclear magnetic resonance (NMR)}
Apart from affirming the intrinsic nature local magnetic moments on Ir$^{5+}$ in this sample, we have performed linear fitting on the NMR mainline-width versus 70 kOe $dc$ magnetic susceptibility ($\chi$) plot (Fig. S10(c)), having temperature as an implicit parameter. The slope of the representative linear fit provides a dipolar coupling constant to be about A = 2.073 $\times$ 10$^{22}$ /cm$^3$-atom. So, it is reasonable to infer that this is due to dipolar interaction fields from the local Ir$^{5+}$ moments and the average distance between the local Ir$^{5+}$ moments and the ${}^{23}$Na nucleus to be about 3.06 {\AA}, further supporting the presence of intrinsic local moments in the SNIO compound.
\begin{figure}
\resizebox{8.6cm}{!}
{\includegraphics[29pt,299pt][502pt,674pt]{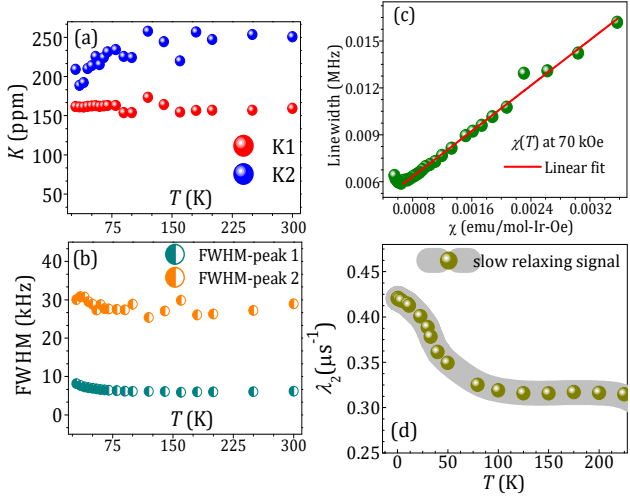}}
\renewcommand{\thefigure}{S\arabic{figure}}
\caption{(colored online) Temperature dependence of NMR-shift (a) and linewidth (b) for the two lines between 30-300 K; (c) Width of the NMR mainline versus $dc$ magnetic susceptibility (measured at 70 kOe) plot (shaded green circles), along with the linear fitting (solid red line); (d) Temperature variation of $\mu$SR slow relaxation rate}
\end{figure}

Finally the ${}^{23}$Na NMR spin-lattice relaxation rate 1/$T_1$ was measured using a saturation recovery sequence. The recovery was found not to be single exponential. Considering that ${}^{23}$Na has $I$ = 3/2 and hence a quadrupole moment, one might expect a multi-exponential recovery in case the full spectrum is not irradiated with the saturating sequence. However, given that the lineshape is a combination of two species of Na with both having quadrupolar effect, we fitted the longitudinal nuclear magnetization recovery to a stretched exponential given by, $M$($t$) = $M_0$[1 - $\exp$(-($t$/$T_1$)$^{\beta}$)]. The relaxation rate 1/$T_1$ (shown in the main panel of Fig. 2(c)) increases with decreasing temperature and exhibits a broad maximum around 150 K. This maximum could not be associated with conventional spin glass freezing, where a critical slowing down of the spin-fluctuations leads to a very short $T_1$ and hence, a diverging spin-lattice relaxation at low and intermediate temperatures~\cite{li2rho3prb-sm,li2rho3prb43-sm,li2rho3prb63-sm}. So, we argue that this arises due to relaxation of the fluctuations arising from the motion of sodium ions. In the Bloembergen-Purcell-Pound formalism (BPP model)~\cite{byionmr37-sm,byionmr38-sm}, the relaxation rate varies as: $\frac{\tau}{(1+\omega^2\tau^2)}$, where, $\tau$ is the correlation time of the Na nuclear dipoles and $\omega$ is the NMR frequency. The correlation time (rate) follows Arrhenius behavior, $\tau$ = $\tau_0$ $\times$ $\exp$($E_A$/$k_B$$T$) where $E_A$ is the activation energy. A maximum occurs around $\omega\tau$ = 1. From the fit to our data (shown by red solid line in the main panel of Fig. 2(c)), we obtain $E_A$/$k_B$ = 400 K and $\tau_0$ = 10$^{-10}$ sec. The ${}^{23}$Na nuclear spin-lattice relaxation rate therefore appears to be dominated in this temperature region by the Na-nuclear dipolar fluctuations and apparently unaffected by Ir-local moment fluctuations. This might be due to the symmetric location of ${}^{23}$Na between the magnetic Ir where the antiferromagnetic fluctuations are filtered out. But, below about 90 K, 1/$T_1$ decreases upon further cooling and displays a power law behavior (inset to Fig. 2(c)) with power 1 in the range $\sim$ 4-30 K (solid red line) and with power 2.5 in the temperature range $\sim$ 30-90 K (solid orange line). This suggests development of magnetic correlations among the Ir-local moments at this low-temperature region, likely in agreement with the $\mu$SR relaxation rate below $\sim$ 100 K (see Fig. 3(b)). Such power-law dependence of 1/$T_1$ could certainly be attributed to the existence of gapless spin excitation spectrum in SNIO, which is in consistent with the spin-orbit 5$d$ iridates Na$_4$Ir$_3$O$_8$~\cite{Mendel-2015-PRL-suppli}, Ba$_3$InIr$_2$O$_9$~\cite{Tushar-PRB2017-sm}, and also other low-dimensional quantum materials~\cite{li2rho3prb-sm,li2rho3prb18-sm,li2rho3prb69-sm}.

At this point, quite surprising is the discrepancy between the nature of NMR spin-lattice relaxation rate $\frac{1}{T_1}$ (inset to Fig. 2(c)) and that of $\mu$SR relaxation rate $\lambda_1^{-1}$ (Fig. 3(b)) possibly due to the difference in their respective characteristic time-windows to which the NMR and $\mu$SR techniques are sensitive. While, the temperature-independent $\lambda_1^{-1}$ in the 30-0.04 K range (Fig. 3(b)) could be assigned to the signature of configurationally degenerate phases with fluctuating order~\cite{Mendel2015PRL-29-sm}, recent local density approximation calculations on the other hand show the possibility of local deformations around the environment of positive charge muons ({\it i.e.,} the muon stopping sites), and hence, affecting the local spin dynamics close to muon probe~\cite{Mendel2015PRL-37-sm}.

\begin{figure}
\resizebox{8.6cm}{!}
{\includegraphics[93pt,368pt][475pt,676pt]{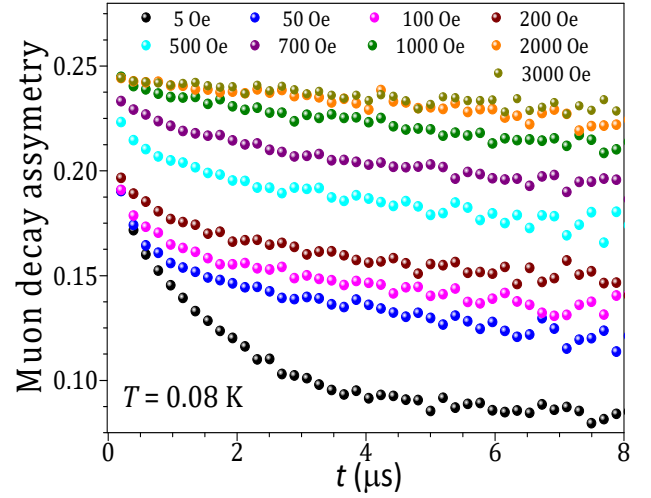}}
\renewcommand{\thefigure}{S\arabic{figure}}
\caption{Evolution of muon polarization at 0.08 K with applied external longitudinal fields}
\end{figure}

\subsection{Heat Capacity}
The measured temperature dependence of the specific heat data ($C_p$ versus $T$) in zero field and several applied high magnetic fields in the low-temperature region are illustrated in the main panel of Fig. S12(a). In absence of any suitable nonmagnetic counterpart, $C_{lattice}$ for Sr$_3$NaIrO$_6$ was extracted after fitting the high-temperature zero-field $C_p$ data ($T$ range, 80-300 K) with Debye-Einstein model, yielding a Debye temperature $\Theta_D$ = 409 K. The high-temperature fitting is then extrapolated down to the lowest measured temperature 2 K (inset to Fig. S12(a)) and taken as $C_{lattice}$ which was subtracted from the total $C_p$. The heat capacity of the sample is thus left out with ($C_M$ + $C_{sch}$).
\begin{figure}
\resizebox{8.6cm}{!}
{\includegraphics[10pt,429pt][564pt,668pt]{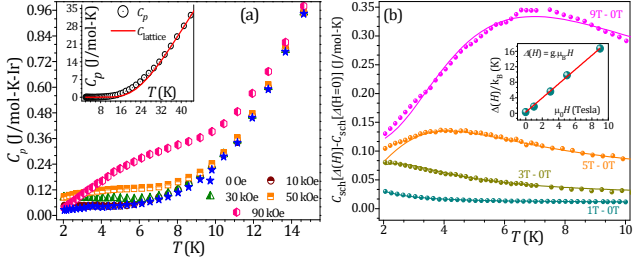}}
\renewcommand{\thefigure}{S\arabic{figure}}
\caption{(a) Temperature variations of the total specific heat $C_p$ in the low-$T$ region for zero field as well as different applied magnetic fields. Inset: Lattice part extraction (solid red line) of the zero field $C_p$ data (open black circles) using Debye-Einstein model; (b) Temperature variations of the [$C_p$($H$) - $C_p$($H$ = 0)] data at the respective applied magnetic fields along with the two level Schottky anomaly fits (solid colored line) in the low-$T$ range; Inset: magnetic field dependent Schottky energy gap ($\frac{\Delta}{k_B}$ versus $H$, shaded dark cyan circles) and the subsequent linear fitting (solid dark red line).}
\end{figure}

In order to obtain the correlated magnetic contribution ($C_M$), the adopted strategy was to remove the Schottky part as: first we subtracted the zero field $C_p$ data, {\it i.e.}, $C_p$($H$=0) from those measured in the applied magnetic fields, {\it i.e.}, $C_p$($H$ $\neq$ 0). Consequently, the temperature variations of this difference ($\Delta$$C_{p,mag}$/$T$) are illustrated in Fig. S12(b), and thereafter has been modeled using,
\begin{equation}
\frac{\Delta C_{p,mag}}{T} = \frac{[C_p(H \neq 0, T) - C_p(H = 0, T)]}{T}
\end{equation}
This is then fitted with two-level Schottky anomaly as
\begin{equation}
\frac{\Delta C_{p,mag}}{T} = \frac{f}{T}[C_{sch}(\Delta(H \neq 0)) - C_{sch}(\Delta(H = 0))]
\end{equation}
where, $f$ represents the percentage of isolated spin centers in the sample; $C_{sch}$($\Delta$) and $C_{sch}$($\Delta_0$) are the Schottky contributions to the specific heat while $\Delta$($H$) represents Zeeman splitting in the applied magnetic field and $\Delta_0$ is energy separation between the two levels at $H$ = 0. The $C_{sch}$($\Delta$) is further defined as,
\begin{equation}
C_{sch}(\Delta) = R(\frac{\Delta}{k_BT})^2\frac{\exp(\Delta/k_BT)}{(1 + \exp(\Delta/k_BT))^2}
\end{equation}
Correspondingly, the two-level Schottky anomaly fits are shown in Fig. S12(b). The fraction of the isolated magnetic centers $f$ accounts to be $\sim$ 5-7 \%, which comes from the antisite-disordered Ir$^{5+}$ magnetic centers, similar to the cases of some $d^5$ iridates~\cite{ba3irti2o9-sm,btrhirprb-sm}. The two-level Schottky splitting gap ($\Delta$/$k_B$) follows a linear magnetic field dependence, $\Delta$ = $g$$\mu_B$$H$ (inset to Fig. S12(b)), as expected for the free spin Schottky anomalies~\cite{ba3irti2o9-sm,btrhirprb-sm,syioprb-sm}. The estimated $g$ value found from this linear fit is about 2.5.



\begin{thebibliography}{99}
\bibitem{MISC1} In field sweep measurements of the spectra at lower frequencies/fields, we could observe a quadrupolar powder pattern with $\nu _Q$ $\approx 0.1$ MHz in addition to the `knee' feature.  In our fixed field measurements where the lineshape is obtained by the Fourier Transform of the echo, the satellites are out of the spectral width.
\bibitem{MISC2} To compare different measurements, we computed the polarization $P(t)=(A(t)-A_b)/(A(t=0)-A_b)$ where $A(t)$ is the measured asymmetry with $A(t=0)\sim 0.28$ and $A_b$ the asymmetry of the muon stopping in the sample holder which was different in the dilution fridge and He flow set-ups.
\end{thebibliography}
\end{document}